# Modeling and Analysis of Three Properties of Mobile Interactive Systems Based on Variable Petri Nets

Ru Yang, Zhijun Ding, *Senior Member*, *IEEE*, Changjun Jiang, MengChu Zhou, *Fellow*, *IEEE*.

***Abstract*—Due to the mobility and frequent disconnections, the correctness of mobile interaction systems, such as mobile robot systems and mobile payment systems, are often difficult to analyze. This paper introduces three critical properties of systems, called system connectivity, interaction soundness and data validity, and presents a related modeling and analysis method, based on a kind of Petri nets called VPN. For a given system, a model including component nets and interaction structure nets is constructed by using VPNs. The component net describes the internal process of each component, while the interaction structure net reflects the dynamic interaction between components. Based on this model, three properties are defined and analyzed. The case study of a practical mobile payment system shows the effectiveness of the proposed method.**

## I. Introduction

In the past few decades, with the continuous promotion and development of (mobile) Internet, many emerging technologies (fields), such as Pervasive Computing, Internet of Thing (IoT), and (Mobile) Cloud Computing were developed and well applied [1]-[5]. They assume that a number of invisible sensing or computational components (entities) interact both with users and with the environment, and can deliver mobile, seamless, transparent, ubiquitous and customized services to users in a context-aware manner. Among the common features of those technologies, mobility and dynamic interactions are critical [6], [32]-[35].

The systems with mobile interactive components (collectively called mobile interactive systems in this paper) have raised many interests. In their execution, components can move and communicate with each other to accomplish a given task, and the movement of components (and environmental change) can lead to frequent and dynamic disconnections among them.

However, due to the mobility and frequent disconnections [2], the correctness of mobile interaction systems are often difficult to guarantee. It is urgent for researchers in this field to understand systems having such characteristics, establish a rigorous mathematical model and provide the proper analysis from different aspects for them.

Formal methods, such as Petri nets (PN) [7], [8], seem to be a good choice to model and analyze those systems. They own rigorous semantics and can describe system features clearly. However, current studies do not give a specific description of the overall structure, component execution, and various interactions of mobile interactive systems, and thus cannot provide a complete modeling and analysis method. This makes them difficult to apply to actual system design. The literature review in this field can be found in the Supplementary file [13]-[31].

We have proposed a new PN, called a Variable Petri Net (VPN) in [40], which can describe the dynamicity of interactions in systems. Based on VPN, this paper focuses on modeling and analysis of properties of mobile interactive systems. Its main contributions are:

1) It introduces a modeling method for a mobile interactive system based on VPN. The modeling process is performed from one component to multi-components by considering its dynamic interactions and contextual changes.

2) It proposes three properties of systems called system connectivity, interaction soundness and data validity, to reflect their execution and interaction conditions, and then presents the analysis methods for these properties.

3) A practical mobile transaction system is modeled and analyzed to show the proposed concepts and methods.

The rest of the paper is organized as follows. The next section introduces mobile interactive systems and a simple example. Section 3 gives the definitions and firing rule of VPN, and also the graphic and behavior analysis techniques for VPN. Section 4 introduces the modeling method for mobile interactive systems by using VPN. Section 5 proposes three properties of systems and their related analysis methods. Section 6 presents a case study. Section 7 discusses the related work. Section 8 concludes this paper.

This work is partially supported by National Natural Science Foundation of China under Grant No. 61173042. (Corresponding author: Zhijun Ding.)

R. Yang, Z. Ding, and C. Jiang are with the Key Laboratory of Embedded System and Service Computing, Ministry of Education, Tongji University, and also with the Department of Computer Science and Technology, Tongji University, Shanghai, 201804, China (e-mail: yangru@tongji.edu.cn, zhijun_ding@outlook.com, cjjiang@tongji.edu.cn).

M. Zhou is with the Institute of Systems Engineering, Macau University of Science and Technology, Macau 999078, China and also with the Department of Electrical and Computer Engineering, New Jersey Institute of Technology, Newark, NJ 07102 USA (e-mail: zhou@njit.edu).

## II. Mobile Interactive Systems And Their Properties

A mobile interactive system is a system consisting of several components that are distributed and can run independently. In its execution, some components are mobile and communicate with each other to accomplish an expected task. Because of the mobility of components and contextual changes, the interactions (connections and disconnections) among components are uncertain and dynamic. Thus system behaviors are very complex. Mobile transaction systems and mobile robot systems belong to this kind of systems.

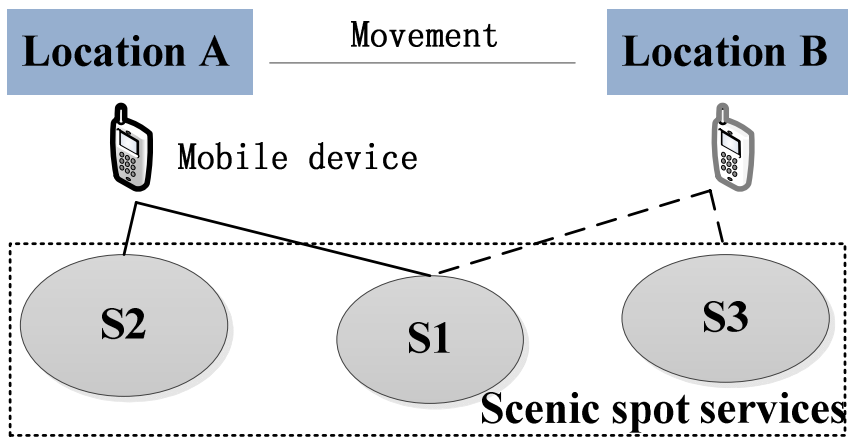

Fig. 1. Example 1 (A tourism system)

Next we introduce a simple example of a mobile interactive system. A tourist attraction designs a Tourist app (system) for the convenience of tourists. Tourists can use mobile phones or other mobile devices (MD) to call (invoke) some services (components) in this system. For simplification, suppose that these exist two scenes A and B, and three Scenic spot services S1, S2 and S3 in the system. S1 is a service for the speciality products reservation that can be used in all locations, and S2 and S3 are services for the voice explanations for A and B. Thus S2 and S3 can be used in locations A and B, respectively. It is noted that there exist connections and disconnections among a mobile device and services. This system is shown in Fig. 1.

Because of the movement of a device, the interactions between it and services are dynamic. Thu some common properties of mobile interaction systems, such as reachability and liveness, are inadequate for them. Some specific properties about the dynamic interactions need to be focused on and verified in the system design stage.

- **System connectivity**

In a mobile interactive system, components can connect and interact with each other. Whether a system has the link (connection) capacity or not is absolutely critical and can determine the connection (interaction) processes among components. Hence, the verification of link capacity of a mobile interactive system is necessary. In this paper we call this property as "**system connectivity**".

- **Interaction soundness**

After the connection or disconnection among components because of their movement or contextual changes, they can begin or suspend to interact with each other. Then an interaction process among them can determine the completion status of a task. Thus if the interactions are executed smoothly and reasonably, they can ensure the correctness and safety of a system. Here we call this property as "**interaction soundness**".

- **Data validity**

Data is an indispensable part in actual systems. There may exist data manipulation in the internal process of a component, and data exchange, transferring or synchronization in the interactions among components in a mobile interactive system. Data interaction may directly influence the result of a system. This can help one to judge whether the parameter for data is valid in an interaction. Here we call this property as "**data validity**".

Then, how to verify these these properties of mobile interaction systems? The modeling and related analysis for systems may be used to answer the above question. And the modeling process should consider both the internal workflow of each component and the interaction process between different components. The VPN model proposed in our previous work [40] is appropriate to model the dynamic interactions in a system, and thus is used as the basic model to construct the system model and perform the analysis of the mobile interactive systems in this work. Then, in the next section, we introduce the basic definitions of VPN [40] and propose some analysis methods of it.

## III. Variable Petri Nets

A Variable Petri Net (VPN) introduces the mapping relations to places and the concept of virtual places to PNs, which allows one to fully describe the mobility and dynamicity in a system. It is an appropriate model to deal with dynamic interactions and contexts in mobile interactive systems [40].

### *A. VPN*

Let $\Sigma$ be a finite set of names which are used to indicate places, tokens and arc weights in the following. Let $\Sigma = V \cup C$, where $V$ is a set of variables, or called formal parameters, and $C$ is a set of constants, or called actual parameters. $C$ contains a special character $\varepsilon$ for the ordinary token by default, and each element $c \in C$ is asscoiated with an arity $n \in \mathbb{N}^+$. For any set $A$, we use $2^A$ to denote all subsets of $A$, $A^*$ to denote all tuples formed by the elements in $A$, and $A^n$ to denote all tuples with exactly $n$ elements of $A$ (whose length is $n$).

In the following, we use $\mathbb{N}$, $\mathbb{N}^+$ and $\mathbb{Z}$ to denote the sets of non-negative integers, positive integers and integers, respectively.

**Definition 2.1 (multiset)**. For any set $A$, a **multiset (bag)** $m$ over $A$ is defined as a mapping $m$: $A \rightarrow \mathbb{N}$. The set of all bags over $A$ is denoted by $\mathbb{N}^A$.

Here we introduce the definition of VPN.

**Definition 2.2 (VPN).** A Variable Petri Net (VPN) $N$ over the universe $\Sigma = C \cup V$ is an 8-tuple $N = (P, T, F, \gamma, W, \varphi, \rho, M_0)$, where

1. $P \subseteq C$ is a set of places. Each place is associated with an arity, which is the length of tuples of tokens in it.
2. $T$ is a finite set of transitions. $P \cap T = \varnothing$.
3. $F \subseteq (P \times T) \cup (T \times P) \cup (T \times V) \cup (V \times T)$ is a set of arcs. Each arc $(t, v) \in (T \times V)$ or $(v, t) \in (V \times T)$ is called a virtual arc, and each variable in a virtual arc is called a virtual place.
4. $\gamma$: $V \rightarrow 2^C$ is a constraint function mapping variable $v \in V$ to a set of constants $X \in 2^C$, and each element $c \in X$ is a place in $P$ or a newly generated place in $C$.
5. $W$: $F \rightarrow \mathbb{N}^{(\Sigma^*)}$ is an arc labelling function (expressions) representing the weight for arcs. Each label can be tuples of constants and variables or the empty set ∅. For any transition $t \in T$, if a variable $v \in V$ meets the condition "$(t, v) \in F$ or $\exists\, e \in \Sigma$: $v \in W(t, e)$", it must also satisfy that "$(v, t) \in F$ or $\exists\, e' \in \Sigma$: $v \in W(e', t)$". For each arc $f \in (T \times P)$ or $(P \times T)$, $W(f) \in \mathbb{N}^{(\Sigma^n)}$, where $n$ is the arity of the place in $f$.

6. $\varphi$: $T \rightarrow \mathbb{B}$ is a guard function associated with each transition, where $\mathbb{B}$ is the set of all Boolean expressions that can be constructed by using constants and variables in Σ.

7. $\rho$: $T \rightarrow (\mathbb{B} \times \Theta)$ is a link function of transitions, where Θ is a series of operations which are done when $\mathbb{B}$ is judged as **true**. For a transition $t$, $\rho(t) = (b, h)$, where $b \in \mathbb{B}$ and $h$ is a do-nothing operation or an operation to add/delete ("+/-") a $\gamma$ constraint to each variable $v$ satisfying that $(t, v) \in F$, denoted as $(v, +/-)$.

8. $M_0$ is an initial marking. A marking of $N$ is a function $M$: $P \rightarrow \mathbb{N}^{(C^*)}$, where $M(p) \subseteq \mathbb{N}^{(C^{n_p})}$ ($n_p$ is the arity of $p$) is the set of tokens residing in $p$.

For a node $x \in P \cup T$, its preset $\bullet x$ and postset $x\bullet$ are subsets of $P \cup T \cup V$ such that $\bullet x = \{y | (y, x) \in F\}$ and $x\bullet = \{y | (x, y) \in F\}$. Let $t$ be a transition. For any $x \in \bullet t$ (or $t\bullet$), if $x \in C$, arc $(x, t)$ is called an input arc (or $(t, x)$ is called an output arc); otherwise, $(x, t)$ is called a virtual input arc (or $(t, x)$ is called a virtual output arc). These arcs are collectively called adjacent arcs of $t$. And $x$ is called the (virtual) pre-place or post-place.

**Definition 3.3.** A guard $\varphi(t)$ of a transition $t$ is a relational expression formed by the constants in $C$ and the variables that exist in the input arc expressions or as the virtual pre-places of $t$.

Next, given $N = (P, T, F, \gamma, W, \varphi, \rho, M_0)$ over $\Sigma = C \cup V$ being a VPN, some definitions about its execution are introduced.

Variables in the VPN can be substituted with constants during its execution. And these substitutions are defined as bindings as follows.

**Definition 3.4 (Binding).** A binding $\beta$ of any transition $t \in T$ is a (partial) function: $V \rightarrow C$ which associates a variable to a constant, and satisfies that if $v \in V$ is a virtual place such that $(t, v)$ or $(v, t) \in F$, the length of tuples in $W(t, v)$ or $W(v, t)$ should be equal to the arity of place $v[\beta]$, or $W(t, v) = \varnothing$.

Binding $\beta$ to virtual pre/post-place $v$, adjacent arc labels and guard function $W(p, t)$, $W(t, p)$, $W(v, t)$, $W(t, v)$ and $\varphi(t)$ of $t$ are denoted by $v[\beta]$, $W(p, t)[\beta]$, $W(t, p)[\beta]$, $W(v[\beta], t)[\beta]$, $W(t, v[\beta])[\beta]$) and $\varphi(t)[\beta]$, respectively.

Using a marking, a place set and $\gamma$, we can uniquely identify a configuration of a changed VPN.

**Definition 3.5 (Configuration of VPN).** Let $M$, $P'$ and $\gamma'$ be a marking, a place set and a constraint function of $N$, then $\Pi = (M, P', \gamma')$ is called a **configuration** Π of $N$. The initial configuration of $N$ is $\Pi_0 = (M_0, P, \gamma)$.

Then the firing rule of VPN is introduced as follows.

**Definition 3.6.** A transition $t \in T$ is enabled in a configuration $\Pi = (M, P, \gamma)$ of $N$ iff there exist one binding $\beta$ satisfying that,

1) $\varphi(t)[\beta] =$ **true**;
2) $\forall p \in P, \forall v \in V$: $(p, t) \in F \Rightarrow M(p) \geq W(p, t)[\beta]$, $(v, t) \in F \Rightarrow (M(v[\beta]) \geq W(v[\beta], t)[\beta]$ and $v[\beta] \in \gamma(v))$;
3) For all variables $v_1, \ldots, v_k$ such that $(v_1, t), \ldots, (v_k, t) \in F$ and $v_1[\beta] = \ldots = v_k[\beta] = p$, $M(p) \geq W(v_1[\beta], t)[\beta] + \ldots + W(v_k[\beta], t)[\beta]$;

which is denoted as $\Pi[t >_\beta$ or $(M, P, \gamma)[t >_\beta$.

**Definition 3.7.** Firing an enabled transition $t$ with binding $\beta$ at a configuration $\Pi = (M, P, \gamma)$ results in the following changes:

1. $P$ into $P'$: for each constant $v[\beta]$ such that $(t, v) \in F$ and $v[\beta] \notin P$, it is added to the place set $P$ ($P = P \cup v[\beta]$), and $M(v[\beta]) = \varnothing$; The final result of $P$ is denoted as $P'$.
2. $\gamma$ into $\gamma'$: for each variable $v$ such that $(t, v) \in F$, if its condition and operation in $\rho(t)$ is $(b, (v, o))$, **then** $\gamma(v) = \gamma(v) \cup \{v[\beta]\}$ if $b[\beta]$ is **true** and $o =$ "+", and $\gamma(v) = \gamma(v) - \{v[\beta]\}$ if $b[\beta]$ is **true** and $o =$ "–"; The final result of $\gamma$ is denoted as $\gamma'$.
3. $M$ into $M'$ such that for each $p \in P$: $M'(p) = M(p) - W(p, t)[\beta] + W(t, p)[\beta] - \sum_{(v,t) \in F: v[\beta]=p} W(v,t)[\beta] + \sum_{(t,v) \in F: v[\beta]=p} W(t,v)[\beta]$.
4. each arc $(v, t)$ or $(t, v) \in F$ in $N$ into solid arc $(v[\beta], t)$ or $(t, v[\beta])$ at the firing of $t$, and then into virtual arc again when a new marking $M'$ is generated.

The new marking $M'$, place set $P'$, and constraint function $\gamma'$ form a new configuration $\Pi' = (M', P', \gamma')$. Thus, Π is transformed to Π′ by firing $t$ with binding $\beta$, represented by $\Pi[t >_\beta \Pi'$ or $M[t >_\beta M'$, $P[t >_\beta P'$ and $\gamma[t >_\beta \gamma'$.

**Definition 3.8.** A sequence of transitions $\sigma = t_1 t_2 \ldots t_k$ is a firing sequence if there exists a series of bindings $\beta = \beta_1 \beta_2 \ldots \beta_k$ and configurations such that $\Pi[t_1 >_{\beta 1} \Pi_1[t_2 >_{\beta 2} \ldots \Pi_{k-1}[t_k >_{\beta k} \Pi_k$, written as $\Pi[\sigma >_\beta \Pi_k$, and $\Pi_k$ is said to be reachable from Π by firing $\sigma$. $\sigma$ can be called a transition (firing) sequence from Π to $\Pi_k$; If Π is reachable from $\Pi_0$, then $R(\Pi)$ is the reachability set of all configurations reachable from Π ($\Pi \in R(\Pi)$).

According to the firing rule, a property called data synchronization of a VPN is defined.

**Definition 3.9 (data synchronization).** The data synchronization of VPN $N$ is defined as: for any variable $v_1 \in V$ and $v_2 \in V$ such that $v_1$ and $v_2$ are contained in two input arc expressions of any transition $t \in T$, if $v_1 = v_2$, then $v_1[\beta] = v_2[\beta]$ at any firing of $t$ with a binding $\beta$.

According to the above definitions, we note that VPN has three main features:

1) Folding of transitions, which makes it has a simpler structure than some existing models, such as CPN;

2) Virtual places and two new functions $\gamma$ and $\rho$, which can reflect the dynamic (dis)connections and context in systems;

3) Configuration including a marking and constraint function $\gamma$, which can represent both state and connection capacity of systems.

This greatly help us describe the dynamic interaction and context in mobile interactive systems. The analysis techniques of VPN are the foundation of the analysis for mobile interactive systems based on VPN. Hence, in the following, the specific analysis techniques for VPNs are proposed.

### *B. Configuration tree generation and Behavior analysis*

Firstly, as the basic analysis technique of PNs, a state space called a configuration tree (CT) of VPN is introduced to reflect some dynamic behaviors and features of systems. Suppose that $N = (P, T, F, \gamma, W, \varphi, \rho, M_0)$ is a VPN over the universe $\Sigma = C \cup V$ where $C$ and $V$ are sets of constant and variable names respectively, and the initial configuration $\Pi_0 = (M_0, \gamma)$ unless otherwise stated in the following discussion.

**Definition 3.10** (**configuration tree).** The configuration tree *CT* of a VPN *N* is a labeled directed tree whose nodes are the reachable configurations (root node is $\Pi_0$) such that there is an arc from configuration $\Pi$ to $\Pi'$ labeled with $(t, \beta)$, satisfying that $t$ is the fired transition, $\beta$ is the used binding, and $\Pi[t>_\beta \Pi'$.

The construction algorithm of CT has been given in our previous work and can be generated by the VPN tool [32]. In CT, all reachable configurations are represented as nodes, and reachability relations among configurations are represented as arcs. Through merging the same nodes into one node, CT is transformed to be a graph (CG). Based on CT and CG, some properties of VPN can be revealed.

Here we just introduce some symbols and new definition about the behaviors of a VPN *N*.

**Definition 3.11 (projection and extension language).** Suppose that $X$ is a finite input alphabet, $Y \subseteq X$. $L_X$ and $L_Y$ are the language on $X$ and $Y$. Let

$$\Gamma_{X\to Y}(L_X) = \{\Gamma_{X\to Y}(\sigma) \in Y^* \mid \forall \sigma \in L_X\},$$
$$\Gamma^{-1}_{X\to Y}(L_Y) = \{\Gamma^{-1}_{Y\to X}(\sigma') \mid \forall \sigma' \in Y^*\},$$

then $\Gamma_{X\to Y}(L_X)$ is called the projection language of $L_X$ from $X$ to $Y$, $\Gamma^{-1}_{X\to Y}(L_Y)$ is called the extension language of $L_Y$ from $Y$ to $X$.

**Definition 3.12 (Control and Data Language).**

$$L_C(N) = \{\sigma \in T^* \mid \Pi_0[\sigma>_\beta \Pi \text{ or } \exists \Pi \in R(\Pi_0): \Pi[\sigma>_\beta \Pi'\}$$
$$L_D(N) = \{\beta \mid \Pi_0[\sigma>_\beta \Pi \text{ or } \exists \Pi \in R(\Pi_0): \Pi[\sigma>_\beta \Pi'\}$$

are called the **control language** and **data** one determined by *N*, respectively.

$L_C(N)$ and $L_D(N)$ are the sets of all possible fired transition sequences and binding sequences from the initial configuration $\Pi_0$ of *N*. The control and data languages of *N* are collectively called languages determined by *N*.

**Definition 3.13.** Given a variable $q \in V$,

$$\mathcal{R}(q) = \{q' \mid \Pi_0[\sigma>_\beta \Pi \wedge \{q \to q'\} \in \beta \wedge q' \in C\}$$

is called $q$'s **mapping set** determined by *N*.

**Definition 3.14.** Given any transition $t \in T$,

$$\mathcal{V}(t) = \{\beta \mid \exists \Pi \in R(\Pi_0): \Pi[t>_\beta \Pi'\}$$

is called the $t$'s **binding function** determined by *N*.

$\mathcal{R}(q)$ is the set of all constants (actual parameters) that can be instantiated from a variable (formal parameter) $q$ with firing all possible fired transition sequences of *N*. $\mathcal{V}(t)$ is the set of all bindings used with firing $t$ in *N*. Both of them can be generated from the language determined by the VPN based on CT.

**Definition 3.15 (connectivity set).** Let *N* be a VPN, $\Gamma(N)$ satisfying that 1) $\gamma \in \Gamma(N)$, and 2) $\Gamma(N) = \{\gamma' \mid \exists \sigma, \beta: (M_0, \gamma)[\sigma>_\beta (M', \gamma')\}$, is called the **connectivity set** of *N*.

The difference between two constraint functions in *N* can be used to denote a new link (binding), broken or unchanged link between formal and actual parameters (variables and constants) in a net's execution. Here we call the set of possible newly created and broken links between formal and actual parameters in the execution of *N* as $\mathbb{C}$ and $\mathbb{K}$, and the set of links that can be sustained as $\mathbb{A}$ of *N*. The set including $\{\mathbb{A}, \mathbb{C}, \mathbb{K}\}$ are called a link set $\mathbb{L}$ of *N*. The algorithm to discover $\mathbb{L}$ can be generated based on CG and $\Gamma(N)$ [40].

Besides the set of links, the sequences of links also need to be considered and analyzed in a VPN. Thus we give a definition.

**Definition 3.16 (connectivity language).** For a firing sequence $\sigma = t_1t_2...t_k$ such that $\Pi[t_1>_{\beta 1}\Pi_1[t_2>_{\beta 2}...\Pi_{k-1}[t_k>_{\beta k}\Pi_k$ and $\Pi_0 = (M_0, \gamma))$, $\Pi_1 = (M_1, \gamma_1)$, … , $\Pi_k = (M_k, \gamma_k)$, the sequence of constraint functions $\gamma, \gamma_1, ..., \gamma_k$ is called its connectivity sequence, the sequence of the difference of them $(\gamma_1-\gamma), (\gamma_2-\gamma_1), ..., (\gamma_k-\gamma_{k-1})$ is called its new-link sequence, and sequence $(\gamma-\gamma_1), (\gamma_1-\gamma_2), ..., (\gamma_{k-1}-\gamma_k)$ is called its broken-link sequence. All possible connectivity, new-link and broken-link sequences from the initial $\gamma$ in *N* are called the connectivity, new-link and broken-link language determined by *N*, denoted by $L_K(N)$, $L_K^N(N)$ and $L_K^B(N)$, respectively.

$L_K(N)$, $L_K^A(N)$ and $L_K^B(N)$ can be generated from CT (CG) of *N*, which are similar to the discovery of $\mathbb{L}$.

## C. *Correlation analysis*

If different VPNs are merged (integrated) into one with the same place or transition, their properties and behaviors may be influenced. This is because merging often reflects some interaction among VPNs (components). Next we discuss its influences on VPNs based on these structures, i.e., the correlation analysis of VPNs.

Then we introduce some rules to judge property and behavior correlations between $N_1$ and $N_2$ with three different merging (interaction) ways.

In the following first two classes, $N_1$ and $N_2$ have no same place and one same transition $t$, and $\varphi(t) = \varphi_1(t) \wedge \varphi_2(t)$. They are merged to *N* by merging *t*.

**(1)** ***Merging by an asynchronous communication transition***

In the condition as shown in Fig. 3(a), $t$ is regarded as a transition transferring data from $N_1$ ($S_1$) to $N_2$ ($S_2$). Here we suppose that ${}^\bullet S_1 = \{t_1\}$, $S_2{}^\bullet = \{t_2\}$, $t_1{}^\bullet - S_1 \neq \emptyset$ and ${}^\bullet t_2 - S_2 \neq \emptyset$. In order to simplify the analysis, the interaction parts ($t_1$, $S_1$, $t$, $S_2$, $t_2$) are ignored in $N_1$ and $N_2$ before merging. Then we have the following results.

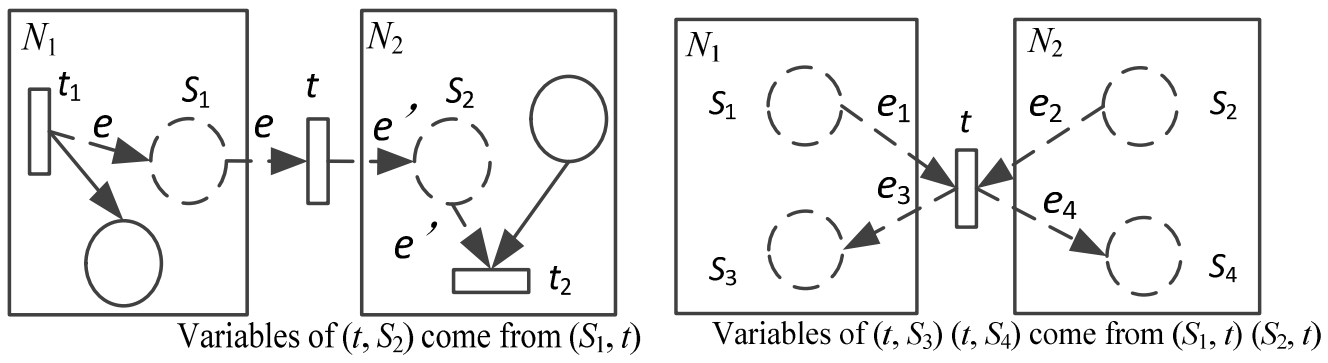

(a) Asynchronization (b) Synchronization

Fig. 3. Two different cases of the merging of two VPNs with a transition

**Theorem 3.1. If** $N_1$ and $N_2$ are live before merging, **then** *N* is live **if** $\varphi(t) = \emptyset$ **or** there exist an infinite number of tokens in some instantiation places of $S_1$ whose assignment to variables in $e$ such that $\varphi(t) =$ **true**; **if** $N_1$ is not live, **then** *N* is not live.

Please see the proofs of Lemma 3.1 and all the following lemmas and theorems in Supplementary File.

**(2)** ***Merging by a synchronous communication transition***

In the condition as shown in Fig. 3(b), $t$ is used for the synchronous communication between $N_1$ and $N_2$. Then the correlation between $N_1$ and $N_2$ can be analyzed as follows.

**Theorem 3.2. If** $N_1$ and $N_2$ are live before merging, **then** $N$ is live **if** $e_1$ and $e_2$ have no same variable, or the binding sequence of each infinite firing sequence of $N_i$ has the same assignment to the same variables of $e_1$ and $e_2$ with the binding sequence of an infinite firing sequence of $N_{3-i}$ ($i$ = 0 or 1).

In the last class, $N_1$ and $N_2$ have no same transition and a same virtual place $S$, and can merge by sharing $S$.

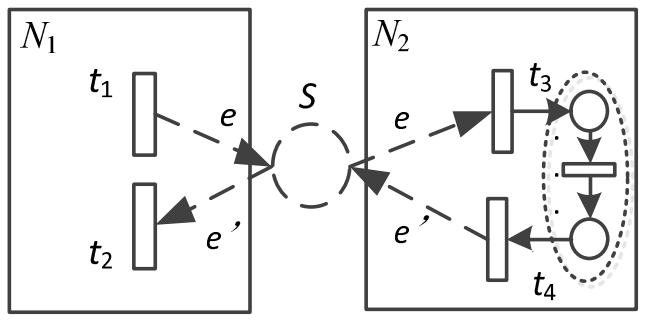

Fig. 4. A case of the merging of two VPNs with a virtual place

**(3) *Merging with a synchronous virtual place* (*interface*)**

As shown in Fig. 4, $N_1$ and $N_2$ both have an input transition and an output transition of $S$. $t_2$ can fire only after $t_1$, and $t_4$ can fire only after $t_3$. This figure shows a Request-Response interaction process between two VPNs. In order to simplify the analysis, the interaction parts ($t_1$, $S$), ($S$, $t_3$), ($t_4$, $S$), ($S$, $t_2$) and $S$ are ignored in $N_1$ and $N_2$ before merging.

In this condition, $S$ is a virtual place and can be instantiated as different places and contain different data (tokens). Thus the possible relations between $N_1$ and $N_2$ become much more complex. The analysis of this kind of merging greatly depends on actual processes of VPNs. Here we give one conclusion.

**Theorem 3.3. If** $N_1$ and $N_2$ are live before merging, **then** $N$ is live **if** the instantiation of each firing of $t_1$ to arc $e$ and $S$ can always make $\varphi(t_3)$ = true meanwhile the instantiation of each firing of $t_4$ to arc $e'$ and $S$ can always make $\varphi(t_2)$ = true.

Based on the definitions and theorems in this section, VPN can be used to analyze mobile interactive systems.

## IV. Modeling of Mobile Interactive Systems

In general, a VPN-based model for a mobile interactive system contains two parts: component and interaction parts. Then a specific modeling method for such systems is introduced next.

### *A. Component nets*

Each kind of components (entities) in a system is modeled as a VPN. Each VPN can be independent or have flow relations with the interaction part. Then we introduce a component net for each component.

**Definition 4.1 (Component net).** The component net (CN) is an VPN defined as $CN$ = ($P$, $T$, $F$, $\gamma$, $W$, $\varphi$, $\rho$, $M_0$) over the universe $\Sigma = C \cup V$, where

(1) $P = P_F \cup P_P \cup P_D \cup P_C \cup P_I$ is the set of places, where $P_F$ contains an initial place and final place, $P_P$ is the set of process places, $P_D$ is the set of data places, $P_C$ is the set of contextual places and $P_I$ is the set of interface places.

(2) $T = T_P \cup T_I$ is the set of transitions, where $T_P$ is the set of internal process transitions and $T_I$ is the set of internal interaction (connection or disconnection) transitions.

It is noted in a CN, there may exist five kinds of internal places: 1) initial place and final place ($P_F$); 2) process places ($P_P$); 3) data (state) places ($P_D$); 4) contextual places ($P_C$); 5) interface places ($P_I$). The initial and final places are used for the start and end of a component. Process places reflect an execution process of a component. Data places store its current state or data, and also its interfaces. The contextual change in it is often denoted by a virtual place that can be instantiated as an actual contextual place. An interface place means the certain interface of a component, which can be an instantiation of a virtual interface. $P_C$ and $P_I$ can be $\varnothing$. There are two kinds of tokens in the places of CNs: black tokens and tuple tokens.

$T$ is the set of transitions, in which $T_P$ models an internal control process (events, actions) in the component, and $T_I$ describes an interaction process of the component with others. The postset of $T_I$ can only be a virtual interface place (variable) or actual interface place.

**Remark 4.1 (context description).** Except an interaction process, the internal process of components, such as the location changes of cars in vehicular cyber physical systems [34], may be uncertain and unknown in the execution. We can regard these conditions as the uncertain contextual changes in systems, and describe them by using a virtual (contextual) place contained in $V$ and also actual contextual place in CN.

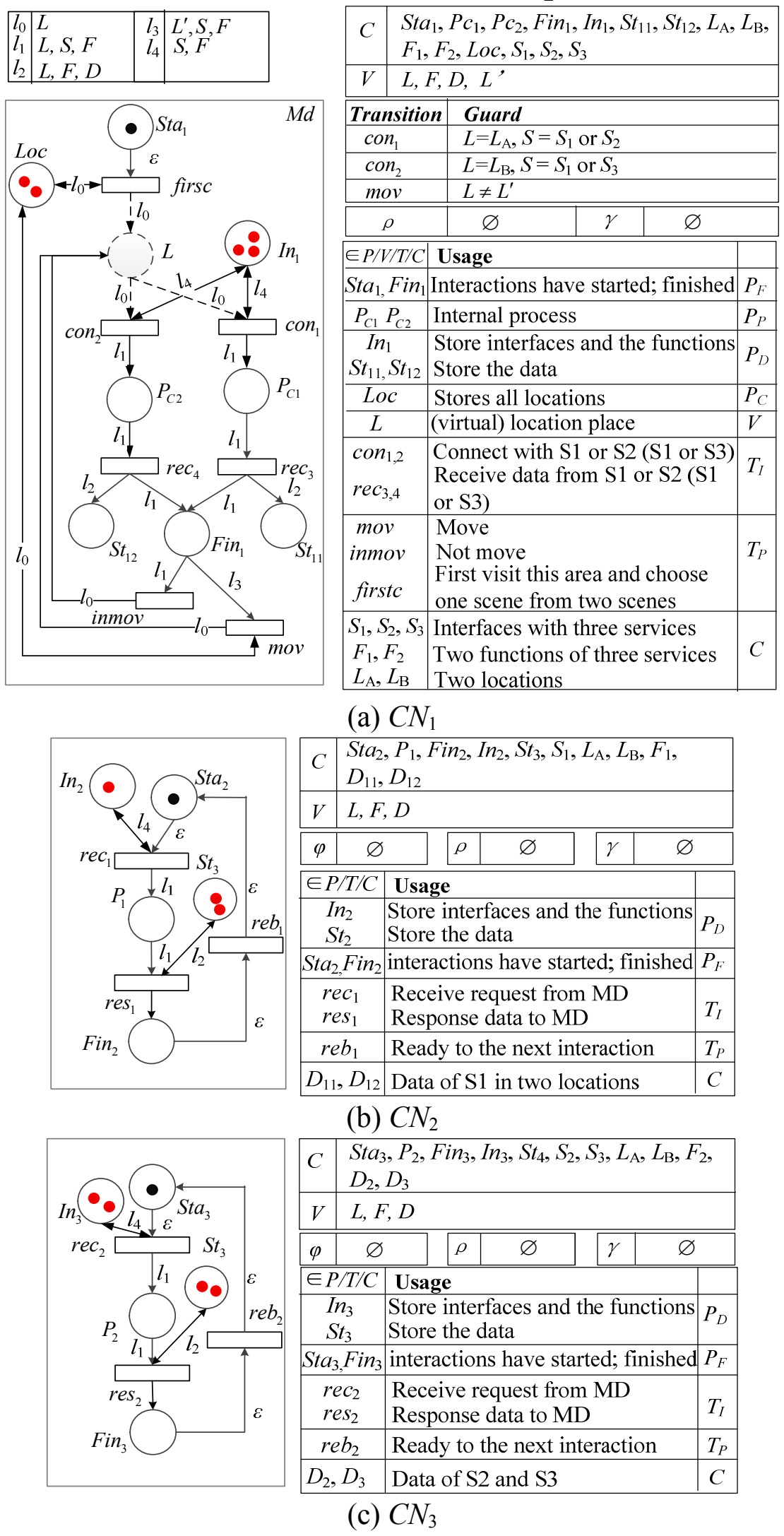

Fig. 5. Component nets (CNs) of Example 1.

In fact, when just considering each component, we may not execute some actions and events (modeled by transitions) if they use the data from other components. Thus the firing rule does not apply to the components directly before they receive data from others.

For example, $CN_1$ for mobile device, $CN_2$ for speciality products reservation service S1, and $CN_3$ for voice explanations services S2 and S3 in Example 1 are constructed based on the above criterion respectively, as shown in Figs. 5(a), (b) and (c). Their initial markings are $M_{10} = \{P_1\{\cdot\}, In_1\{\{S_1, F_1\}, \{S_2, F_2\}, \{S_3, F_2\}\}, Loc\{L_A, L_B\}\}$, $M_{20} = \{P_2\{\cdot\}, In_2\{S_1, F_1\}, St_3\{\{L_A, F_1, D_{11}\}, \{L_B, F_1, D_{12}\}\}\}$ and $M_{30} = \{P_4\{\cdot\}, In_3\{\{S_2, F_2\}, \{S_3, F_2\}\}, St_3\{\{L_A, F_2, D_2\}, \{L_B, F_2, D_3\}\}\}$, respectively. The parameters of each CN, and the usage and classification of (virtual) places and transitions in it are also given. In addition, it is noted that there exist another virtual place $L$ to denote the location in $CN_1$. Its instantiation to contextual place $L_A$ or $L_B$ can reflect the contextual changes, i.e., changes of locations of a mobile device.

### B. *Interaction structure net*

The possible interactions among components in mobile interaction systems are dynamic and may be determined during component execution. Thus the interaction part is modeled by using some virtual places and transitions in VPN, which is different from fixed structures in other PN models. All possible connected interfaces in a system are considered as a predefined set called Interface set $\mathbb{I} \subset C$ ($C$ is the constant set). We can next model the interactions by using interaction structure nets among components.

**Definition 4.2 (Interaction structure net).** Suppose that components $CN_{i1}$-$CN_{ij}$ have interaction relations. An interaction structure net (ISN) among them is a VPN defined as $ISN = (P^I, T^I, F^I, \gamma^I, W^I, \varphi^I, \rho^I, M^I_0)$ over the universe $\Sigma^I = C^I \cup V^I$, where

(1) $P^I$ is a set of interface places.

(2) $T^I = T_{II} \cup T_{EI}$ is the set of interaction transitions. $T_{II}$ is a set of internal interaction transitions in $CN_{i1}$-$CN_{ij}$, and $T_{EI}$ is a set of external interaction transitions.

$P^I$ is a set of interface places, which denotes the channels among components.

$T^I$ is a set of interaction transitions, which represents message sending or receiving. $T_{EI}$ can be $\varnothing$.

Specific modeling details are described as follows.

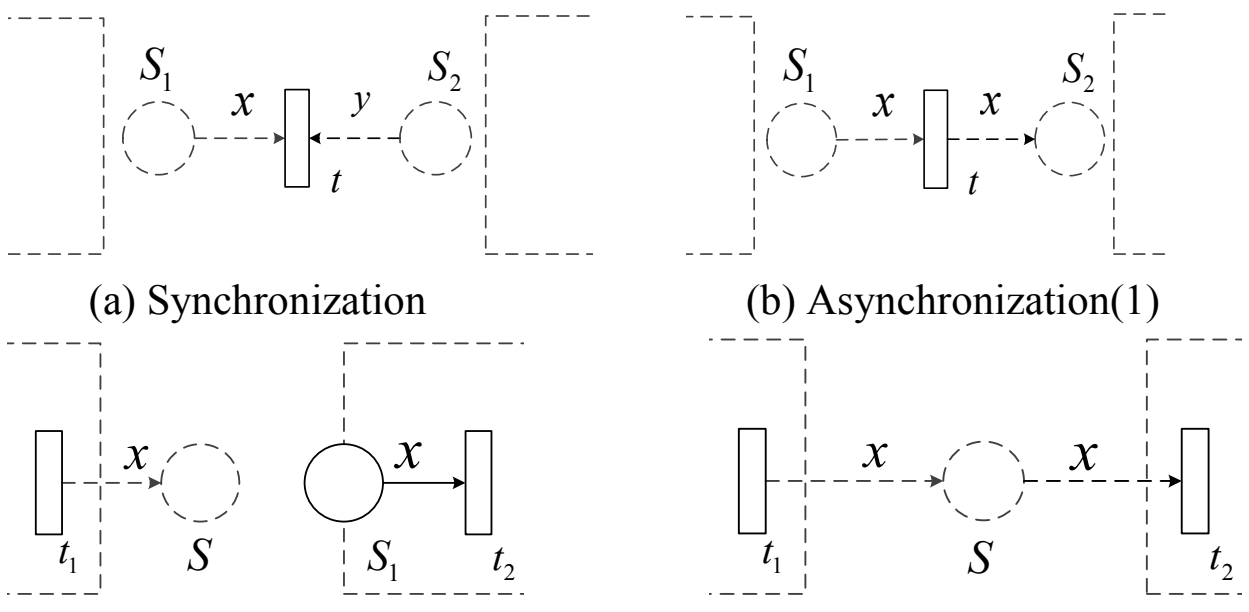

Fig. 6. Models for the interaction part.

● Dynamic interactions among components.

1) Synchronization. A synchronous interaction between two components are modeled as an interaction transition and a couple of virtual interfaces (places) in $V^I$, as shown in Fig. 5(a).

2) Asynchronization. An asynchronous interaction between two components can be modeled in three ways as shown in Figs. 5(b), (c) and (d): one is an external interaction transition and a couple of virtual interfaces, which reflect a detailed interaction process; one is a virtual interface, several actual interfaces and internal interaction transitions, which reflect the abstraction of the interfaces of a component interaction process; the last one is one virtual interface (and several internal interaction transitions in components) for simplicity. The virtual interface (place) can be instantiated into different (existing or new created) actual interfaces in net execution.

● The predefined (possible) connections & disconnections (failures) of components. On one hand, if it has been already known that some components have connected and thus some actual interfaces (places) of them can certainly be instantiated from some virtual interfaces (places), these relations between actual and virtual interfaces are added in the initial $\gamma$ function; On the other hand, if a component can connect or disconnect with others by an action (event), an interaction transition with a function $\rho$ is added as the input transition of a virtual place in a VPN model for the action. $\rho$ is defined to bind or unbind the relation between the virtual place (interface) and the interface of the component.

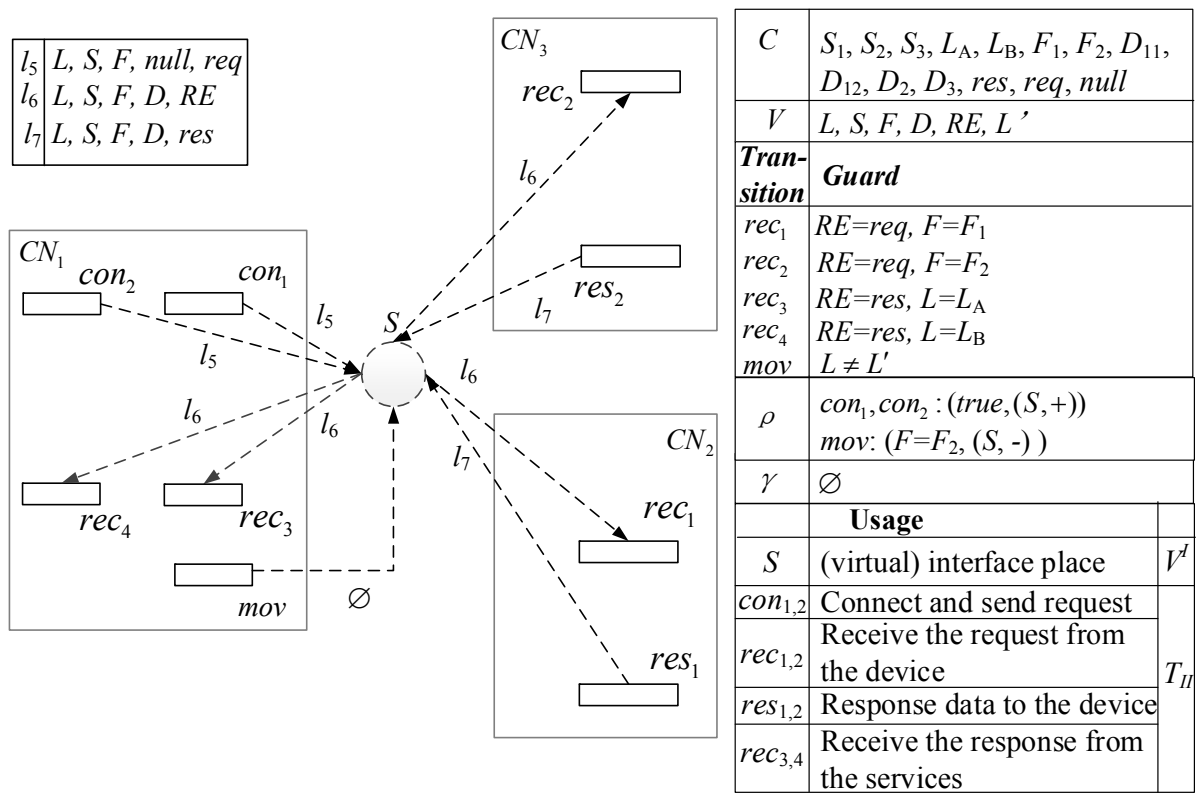

Fig. 7. The interaction structure net (*ISN*) of Example 1.

*ISN* for Example 1 is constructed accordingly as shown in Fig. 7. $\mathbb{I} = \{S_1, S_2, S_3\}$. The interface is modeled by one virtual place $S$, and $S$ can be bound with the interface of services and also unbound with some of them because of the mobility. The information (message) transferred among components are denoted as a tuple $(L, S, F, D, RE)$, where $L$ stands for a location, $S$ for the interface of a service, $F$ for the function of a service, $D$ for the data, and $RE$ for a message type (*res* or *req*).

### C. *Multi-component net*

Finally, we combine component nets and interaction structure nets, and introduce a multi-component net that can describe component execution and mutual interaction among components.

**Definition 4.3 (Multi-component net)**

A multi-component net (MCN) is a VPN defined as $N^m = (\Omega, \Xi)$ over the universe $\Sigma$, where

(1) $\Omega = \{CN_1, CN_2 \ldots CN_m\}$ is a series of CNs;

(2) $\Xi = \{ISN_1, ISN_2 \ldots ISN_n\}$ is a set of ISNs among CNs;

(3) $N^m.\ \Sigma\ (P, T, F, \gamma, W, \varphi, \rho, M_0) = \Omega.\ \Sigma\ (P, T, F, \gamma, W, \varphi, \rho,$

$M_0) \cup \Xi$. $\Sigma$ $(P, T, F, \gamma, W, \varphi, \rho, M_0)$.

MCN contains several component nets and interaction structure nets, which are all VPNs. It follows the firing rule of VPN. It is variable, and can describe concurrent components and dynamic interactions among them.

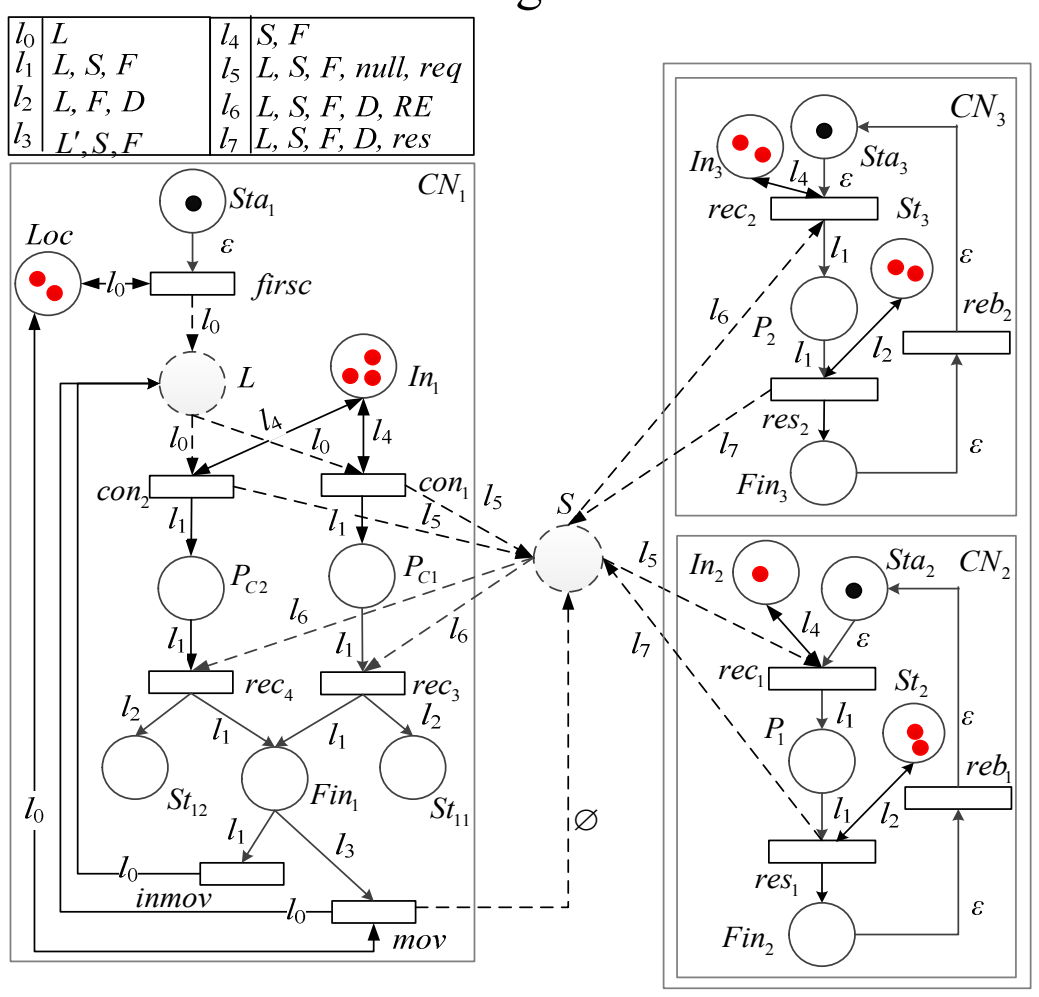

(a) VPN model

| C | $Sta_{1\text{-}3}$, $Fin_{1\text{-}3}$, $P_{1\text{-}2}$, $P_{c1\text{-}2}$, $In_{1\text{-}3}$, $St_{11\text{-}12}$, $St_{2\text{-}3}$, $S_{1\text{-}3}$, $L_A$, $L_B$, $F_{1\text{-}2}$, $Loc$, $D_{11\text{-}12}$, $D_2$, $D_3$, $res$, $req$, $null$ |
|---|---|
| V | $L, S, F, D, RE, L'$ |
| ρ | $con_1, con_2$ : $(true, (S, +))$<br>$mov$: $(F=F_2, (S, -))$ |
| γ | ∅ |

| Transition | Guard |
|---|---|
| $con_1$ | $L=L_A$, $S=S_1$ or $S_2$ |
| $con_2$ | $L=L_B$, $S=S_1$ or $S_3$ |
| $rec_1$ | $RE=req$, $F=F_1$ |
| $rec_2$ | $RE=req$, $F=F_2$ |
| $rec_3$ | $RE=res$, $L=L_A$ |
| $rec_4$ | $RE=res$, $L=L_B$ |
| $mov$ | $L \neq L'$ |

(b) Parameters and meanings of places and transitions in it

Fig. 8. VPN model $N^m{}_{e1}$ for Example 1

Data synchronization (matching) is realized by the same formal parameter on multiple input arcs of an interaction transition in ISN or internal transition in CN.

The VPN model $N^m{}_{e1}$ for Example 1 is constructed as shown in Fig. 8. $N^m{}_{e1} = (\Omega_1 = \{CN_1, CN_2, CN_3\}, \Xi_1 = \{ISN\})$. And its initial marking is $M_0 = M_{10} \cup M_{20} \cup M_{30} = \{P_1\{\cdot\}, In_1\{\{S_1, F_1\}, \{S_2, F_2\}, \{S_3, F_2\}\}, Loc\{L_A, L_B\}, \{P_2\{\cdot\}, In_2\{S_1, F_1\}, St_3\{\{L_A, F_1, D_{11}\}, \{L_B, F_1, D_{12}\}\}, \{P_4\{\cdot\}, In_3\{\{S_2, F_2\}, \{S_3, F_2\}\}, St_3\{\{L_A, F_2, D_2\}, \{L_B, F_2, D_3\}\}\}$. It is noted that Fig. 8 gives a description for the system in Example 1 and can be used to answer the questions if the modeled system has good properties.

Then in order to give the targeted analysis for a mobile interaction system based on a VPN model, we introduce three properties and their related analysis methods next.

## V. Analysis of Mobile Interactive Systems Based on VPNs

In this section, we introduce the definitions and verification of properties based on the VPN models. Suppose that $N^m = (P, T, F, \gamma, W, \varphi, \rho, M_0)$ over the universe $\Sigma = C \cup V$ is a VPN for a mobile interactive system, and the initial configuration $\Pi_0 = (M_0, \gamma)$. The interface set is $\mathbb{I}$.

### A. Three properties

Then we introduce the following properties of $N^m$.

**(1) System connectivity**

**Definition 5.1 (system connectivity)**. The system connectivity is defined as a system that has the link capacity, i.e., the mapping set of interface $I$ of $N^m$ is not null ($\mathcal{R}(I) \neq \varnothing$).

For example, in Example 1, the system connectivity can be analyzed by judging $\mathcal{R}(S) = \{S_1, S_2, S_3\} \neq \varnothing$. Hence, Example 1 owns "system connectivity".

**(2) Interaction soundness**

**Definition 5.2 (interaction soundness).** The **interaction soundness** in a system is defined as:

- Interaction can be finished, i.e., final places can have tokens.
- All possible created, connected and disconnected links (reflected by the binding/unbinding relations between some virtual and actual places in $\gamma$ functions) of the system are not beyond the link capacity of the system. That is, $\mathcal{I} \subset \mathbb{I}$ **where** $\mathcal{I}$ is a set of actual interfaces (constants) in $\mathbb{L}$ of $N^m$.
- The interfaces can be used to transfer data after its connection and become unavailable after disconnection (also called as having good usability), i.e., for each constant $I_A$ of $N^m$ such that $(I, I_A) \in \mathbb{A} / \mathbb{C}$ (or $\mathbb{K}$) and $I$ is a variable, $I_A \in \mathcal{V}(t)$ (or $\notin \mathcal{V}(t)$) where $t$ is a (or any) interaction transition.

For example, the interaction soundness of $N^m{}_{e1}$ in Fig. 8 is analyzed as follows. Firstly, note that all final places can obtain tokens; Secondly, $\mathbb{L} = \{\mathbb{A} = \{S \to S_1\}, \mathbb{C} = \{S \to \{S_1, S_2, S_3\}\}, \mathbb{K} = \{S \to \{S_2, S_3\}\}\}$ can be computed, and thus $\mathcal{I} = \{S_1, S_2, S_3\} \subset \mathbb{I}$; Thirdly, it is found that interfaces among $CN_1$ (mobile device), $CN_2$ (Service S1) and $CN_3$ (Services S2 and S3) have good usability: $S_1$, $S_2$ and $S_3$ are connected after the firing of transitions $con_{1\text{-}2}$ and can transfer data by using the interface $S_1$, $S_2$ and $S_3$ through firing transitions $rec_{1\text{-}4}$ and $res_{1\text{-}2}$, and $S_2$ or $S_3$ become disconnected by firing transition $mov$ and meanwhile $S_2$ and $S_3$ become unavailable interface. In summary, the system of Example 1 has interaction soundness.

**(3) Data validity**

**Definition 5.3 (data validity).** Data validity in a system is defined as:

- Data synchronization is satisfied (Definition 3.6).
- Data interaction has the non-repeatability and atomicity, i.e., before the data in an actual interface sent by a component is transferred to the receivers (or be discarded), this component cannot send another data to this actual interface (Before tokens generated in any actual interface $I_A$ by firing a transition $t$ such that $(t, I_A) \in F$ are consumed (transferred) by a transition $t'$ such that $(I_A, t') \in F$, $t$ cannot be fired).
- There exist no invalid data that cannot be received by components or be discarded, i.e., any token in the interface $I$ can always be consumed (transferred) by firing an interaction transition.

### B. Analyzing algorithms

The analysis of the first property "system connectivity" can be performed by a set of the VPN model. Thus here we give two algorithms to analyze the latter two properties.

**Algorithm 5.1. Interaction Soundness Analysis**

Input: A VPN $N^m$, its CT and interface set $\mathbb{I}$

Output: sound or not

1) **If** the configurations (markings) representing the final places have the tokens are reachable in CT of the net, **then**

2) Obtain $\mathbb{L} = \{\mathbb{A}, \mathbb{C}, \mathbb{K}\}$ of $N^m$, where $\mathbb{C}$ (new adding part of $\gamma$ functions) reflects the newly created connections, $\mathbb{K}$ (deleting part of $\gamma$ functions) reflects the disconnections, and $\mathbb{A}$ (terms not changed in $\gamma$ functions) reflects the sustained connections without broken.

3) **If** $\mathcal{I} \subset \mathbb{I}$ such that $\mathcal{I}$ is set of actual interfaces (constants) in $\mathbb{L}$ **, then**

4) **If** after any relation $\{I \rightarrow i\}$ (in $\mathbb{C}$ or $\mathbb{A}$ ) between a virtual input place $I$ and a constant $i$ is added in the $\gamma$ function by firing interaction transitions, $I$ can be instantiated as the available interface $i$ to send data in the execution, i.e., $\{I \rightarrow i\}$ is contained in $\mathcal{V}$ ($t$) of an interaction transition $t$, **then**

5) **If** after any relation $\{I \rightarrow o\}$ (in $\mathbb{K}$) between a virtual place $I$ and a constant $o$ is not contained in $\gamma$ functions, $I$ never can be instantiated as the unavailable interface $o$ to send data in the execution before $o$ becomes available again, i.e., $\{I \rightarrow o\}$ is not in the binding $\beta$ of any transition after the disconnection, then

6) **return true;**

7) **return false;**

8) **End.**

Algorithm 5.1 corresponds to Definition 5.2. Step 4 is used to verify the connections while Step 5 verifies the disconnections in mobile interactive systems.

**Algorithm 5.2. Data Validity Analysis.**

Input: A VPN $N^m$, its CT

Output: valid or not

1) **If** there is only one instantiation of each formal parameter (for the data) in the binding $\beta$ of one firing of any transition according to Definition 3.6, **then**

2) Analyze if different instantiations of this parameter will lead to errors.

3) **If** for each actual interface place $I_A$, if $M_0[\sigma t > M$ where $\sigma$ is a firing sequence without transition $t$, $(t, I_A) \in F$ and $M(I_A) \neq \varnothing$, then any firing sequence $\sigma'$ such that $M[\sigma'>$ cannot contain $t$ unless it contain a transition $t'$ such that $(I_A, t') \in F$, **then**

4) **If** for each marking $M$ such that $M(I) \neq \varnothing$, there exists a transition $t$ such that $M[t>$, **then**

5) **return true;**

6) **return false;**

7) **End.**

Algorithm 5.2 corresponds to Definition 5.3 and aims to analyze the data validity in mobile interactive systems.

## VI. CASE STUDY

### A. System description

In this section, we use a **tourist system** as Example 2, which is extended from a practical example in [11]-[12], to demonstrate the proposed concepts and methods.

Three components: Client (with mobile device), Merchant (Tourism app) and TPP (third-party platform) participate in an execution process of this system. The tourist system has two functions: ticket-buying ($f1$) and hotel reservation ($f2$). The hotel reservation function needs the interaction between Client and Merchant while the ticket-buying function needs the interaction among three components. A mobile device can move from one place to another, which leads to fragile connections between it and Merchant or TPP. The schematic diagram is shown in Fig. 9(a), and the more detailed interaction process among them are shown in Figs. 9(b) and (c).

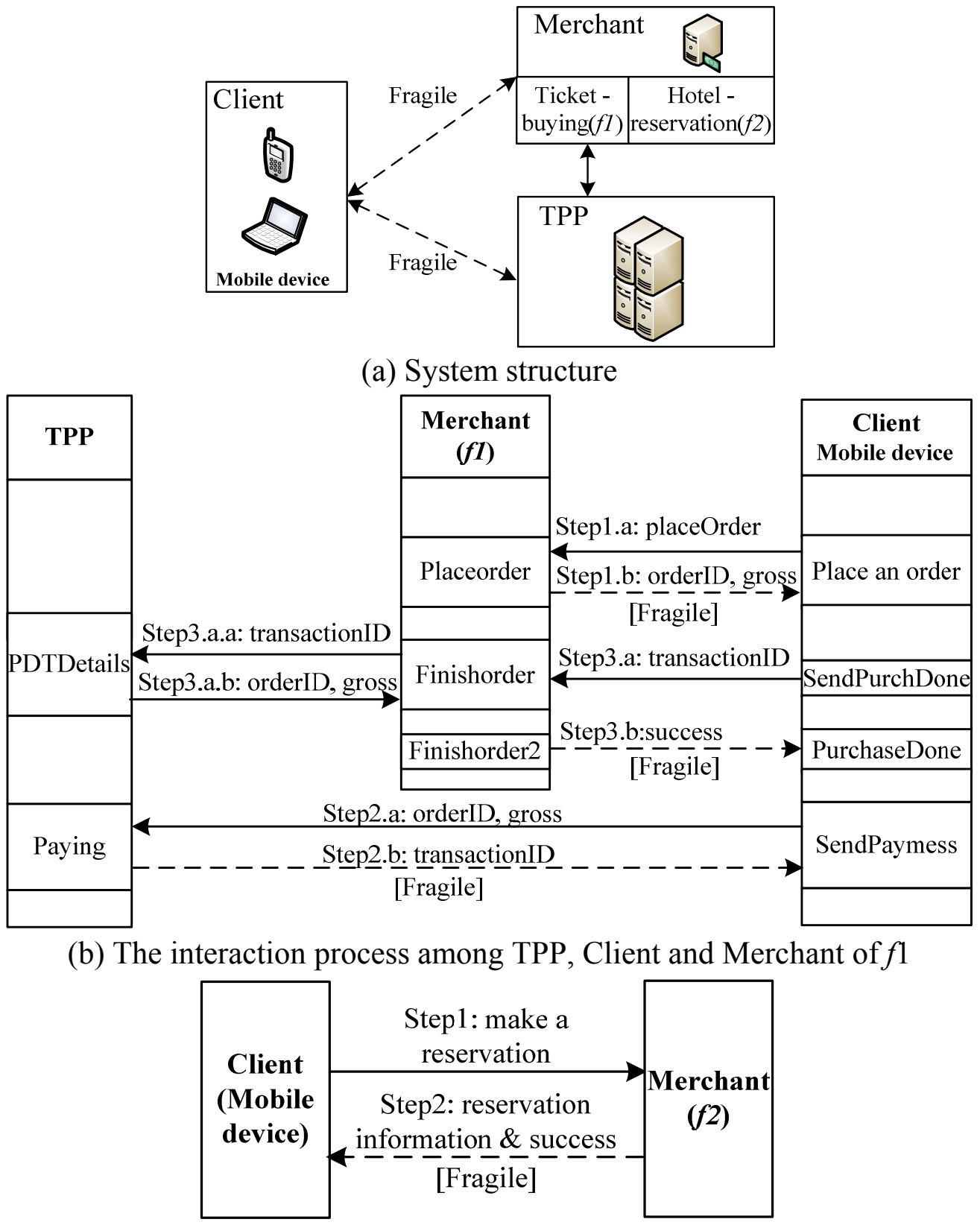

(a) System structure

(b) The interaction process among TPP, Client and Merchant of $f1$

(c) The interaction process between Client and Merchant of $f2$

Fig. 9. An extended tourist system (Example 2)

Fig. 9(b) shows the running process of function $f1$ [11], [12]. Firstly, Client can place an order and send Step1.a to invoke the API-placeorder of Merchant, which inserts the order information including (orderID, gross) to the data storage. Since the order is unpaid, the status is set to *pending*. Then Merchant responds with Step 1.b to transmit the order information to Client and redirects its browser to TPP, where Client pays according to the order information. TPP can record the payment details and return transactionID for the payment via Step 2.b. Client invokes API-finishOrder of Merchant in Step3.a after the payment to finalize the invoice. Further, Merchant makes a call to API-PDTDetails of TPP in Step3.a.a by using transactionID to get the payment details through Step3.a.b. Based on OrderID in the payment details, it finds the order from its data storage. Once the order is located and its status is found to be *pending*, change the status from *pending* to *paid* and a confirmation is finally sent to Client in Step3.b.

Fig. 9(c) shows the running process of function $f2$. It is similar to a part of the process in Fig. 9(b). After receiving a reservation request of Client, Merchant creates a reservation and changes its state from *unres* to *res*, and returns the reservation information to Client.

In order to insure the correctness of component execution and interaction processes of this system, a system model is needed. VPN with an unfixed structure is appropriate. Hence in the following, it is used to model and analyze this system.

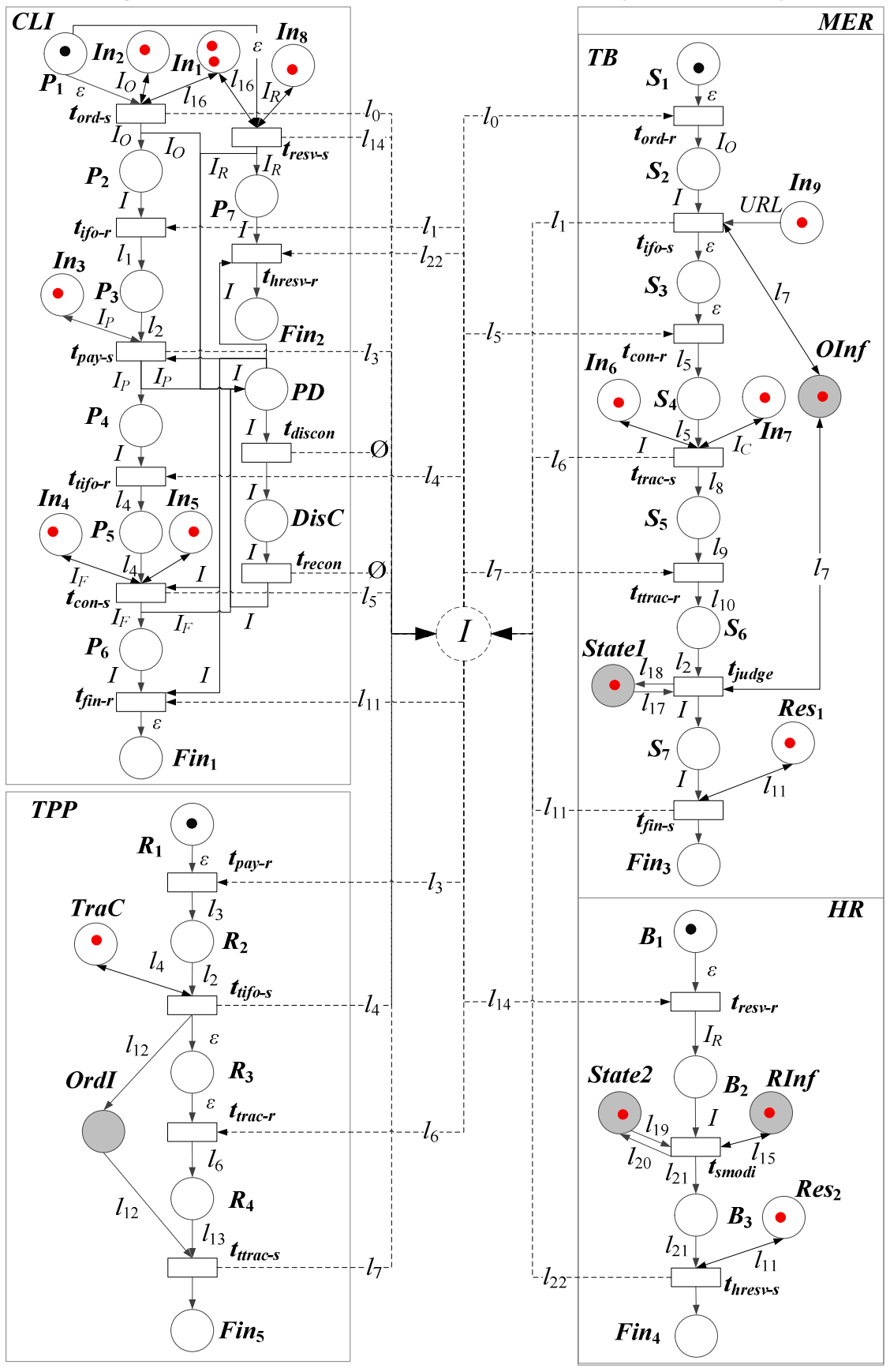

| Arc | Expression (weight) | Arc | Expression (weight) |
|---|---|---|---|
| $l_0$ | $F$, $I_O$ | $l_{12}$ | *TransID, OrderID, Gross* |
| $l_1$ | *OrderID, Gross, URL* | $l_{13}$ | *TransID*, $I$ |
| $l_2$ | *OrderID, Gross*, $I$ | $l_{14}$ | $F$, $I_R$ |
| $l_3$ | *OrderID, Gross*, $I_P$ | $l_{15}$ | *ReservIfo* |
| $l_4$ | *TransID* | $l_{16}$ | $F$, $I$ |
| $l_5$ | *TransID*, $I_F$ | $l_{17}$ | *OrderID, Gross, Status* |
| $l_6$ | *TransID*, $I_C$ | $l_{18}$ | *OrderID, Gross, paid* |
| $l_7$ | *OrderID, Gross* | $l_{19}$ | *ReservIfo, Status* |
| $l_8$ | $I_F$, $I_C$ | $l_{20}$ | *ReservIfo, res* |
| $l_9$ | $I_F$, $I$ | $l_{21}$ | *ReservIfo*, $I$ |
| $l_{10}$ | *OrderID, Gross*, $I_F$ | $l_{22}$ | *ReservIfo, Result* |
| $l_{11}$ | *Result* | | |

| Transition | Guard function |
|---|---|
| $t_{ord\text{-}s}$ | $I = placeorder_{C\text{-}M}$, $F = f1$ |
| $t_{resv\text{-}s}$ | $I = reservation_{C\text{-}M}$, $F = f2$ |
| $t_{ord\text{-}r}$ | $I = placeorder_{C\text{-}M}$, $F = f1$ |
| $t_{con\text{-}r}$ | $I = transconfirm_{C\text{-}M}$ |
| $t_{pay\text{-}r}$ | $I = transaction_{C\text{-}T}$ |
| $t_{trac\text{-}r}$ | $I = transconfirm_{M\text{-}T}$ |
| $t_{resv\text{-}r}$ | $I = reservation_{C\text{-}M}$, $F = f2$ |
| $t_{fin\text{-}r}$ | *Result = success* |
| $t_{hresv\text{-}r}$ | *Result = success* |
| $t_{judge}$ | *Status = pending* |
| $t_{smodi}$ | *Status = unres* |
| $\gamma$ | $\varnothing$ |
| $\rho$ | $t_{dis\text{-}con}$ : if $\{I \rightarrow M(PD)\} \in \gamma$, then - <br> $t_{recon}$ : + |

(a) VPN model $N^m e_2$

| Transtion | Meaning | |
|---|---|---|
| $t_{ord\text{-}s}$, $t_{ord\text{-}r}$ | Step1.a | $T_I$ |
| $t_{ifo\text{-}s}$, $t_{ifo\text{-}r}$ | Step1.b | |
| $t_{pay\text{-}s}$, $t_{pay\text{-}r}$ | Step2.a | |
| $t_{tifo\text{-}s}$, $t_{tifo\text{-}r}$ | Step2.b | |
| $t_{con\text{-}s}$, $t_{con\text{-}r}$ | Step3.a | |
| $t_{trac\text{-}s}$, $t_{trac\text{-}r}$ | Step3.a.a | |
| $t_{ttrac\text{-}s}$, $t_{ttrac\text{-}r}$ | Step3.a.b | |
| $t_{fin\text{-}s}$, $t_{fin\text{-}r}$ | Step3.b | |
| $t_{resv\text{-}s}$, $t_{resv\text{-}r}$ | Reserve | |
| $t_{con}$ <br> $t_{discon}$ | ReConnect <br> Disconnect | |
| $t_{hresv\text{-}s}$, $t_{hresv\text{-}r}$ | Return a reservation result | |
| $t_{judge}$; $t_{smodi}$ | Verify the order status;modify the reservation status | $T_P$ |

| Place | Meaning | |
|---|---|---|
| $In_1$-$In_9$ | Interface storage places | $P_D$ |
| *Oinf, OrdI* | Order information storage places | |
| *TraC* | Transaction information storage place | |
| $Res_{1,2}$ | Result places | |
| *RInf* | Reservation information storage place | |
| *PD* | Place storing interfaces possible to disconnect | |
| *DisC* | Place storing interfaces which have disconnected | |
| $State_{1,2}$ | Status places for the order and reservation in *f1* and *f2* | |
| $Fin_1$-$Fin_5$ <br> $P_1$,$R_1$,$S_1$,$B_1$ | Finish places <br> Initial places | $P_F$ |
| *I* | Virtual interface place | $V^I$ |

(b) Meanings of transitions and places

Fig. 10. VPN model $N^m{}_{e2}$ for Example 2.

### B. *Modeling process*

Components in the system are modeled as three CNs, i.e., *CLI*, *MER* and *TPP*. *MER* contains two parts *TB* and *HR* for ticket-buying and hotel-reservation. There is one interaction structure net *ISN* including a virtual place and several internal interaction transitions in CNs. Then the VPN model $N^m{}_{e2}$ for Example 2 is shown in Fig. 10.

$N^m{}_{e2} = (\Omega_2 = \{CLI, MER, TPP\}, \Xi_2 = \{ISN\}) = (P, T, F, \gamma, W, \varphi, \rho, M_0)$ under $\Sigma$, where,

1) $P$, $T$, $F$, $\gamma$, $W$, $\varphi$ and $\rho$ are given in Fig. 10, and $M_0 = \{P_1\{\cdot\}$, $In_1\{\{placeorder_{C\text{-}M}, f1\}, \{reservation_{C\text{-}M}, f2\}\}$, $In_2\{orderifo_{C\text{-}M}\}$, $In_3\{transactionifo_{C\text{-}T}\}$, $In_4\{final1_{C\text{-}M}\}$, $In_5\{transconfirm_{C\text{-}M}\}$, $In_6\{transconfirm_{M\text{-}T}\}$, $In_7\{confirmifo_{M\text{-}T}\}$, $In_8\{final2_{C\text{-}M}\}$, $In_9\{transaction_{C\text{-}T}\}$, $OInf\{orderid, gross\}$, $TraC\{transactionid\}$, $State1\{orderid, gross, pending\}$, $State2\{reservationifo, unres\}$, $Res_1\{success\}$, $Res_2\{success\}$, $RInf\{reservationifo\}$, $S_1\{\cdot\}$, $S_8\{\cdot\}$, $R_1\{\cdot\}\}$.

(2) $\Sigma = C \cup V$ where $C = \{P_1\text{-}P_6, S_1\text{-}S_7, R_1\text{-}R_4, B_1\text{-}B_3, In_1\text{-}In_9, Fin_1\text{-}Fin_5, OInf, TraC, Res_1, Res_2, RInf, OrdI, PD, DisC, f_1, f_2, placeorder_{C\text{-}M}, orderifo_{C\text{-}M}, transaction_{C\text{-}T}, transactionifo_{C\text{-}T}, transconfirm_{C\text{-}M}, transconfirm_{M\text{-}T}, confirmifo_{M\text{-}T}, final1_{C\text{-}M}, reservation_{C\text{-}M}, final2_{C\text{-}M}, orderid, gross, pending, paid, transactionid, success, res, unres, reservationifo\}$; $V = \{F, I, I_O, I_P, I_F, I_R, I_C, TransID, OrderID, Gross, Status, URL, ReservIfo, Result\}$.

Fig. 10(b) shows the usage of several constants (places) and transitions in $N^m{}_{e2}$. There exist one virtual place $I$ for the connection between Client and Merchant or Client and TPP or Merchant and TPP. New places (interfaces) can be generated by the instantiations of variable $I$ when components interact. Thus the dynamicity and mobility of this system can be directly and vividly described.

### C. *Analysis process*

We have developed a tool called a VPN tool. It can be used to draw VPN, generate its CT and then give some analysis results [40]. Here we give the part of CT of $N^m{}_{e2}$ and some results based on CT in Fig. 11. In Fig. 11(a), there exist four complete paths. PATH 1 (2) means that function *f1* is executed among the components without (with) disconnections of three interfaces. PATH 3 (4) means that function *f2* is executed between components without (with one) disconnection. Then we analyze three fundamental properties and two other requirements for this system by using the methods presented in the last section.

(1) **System connectivity**

The mapping set of $I$ satisfies $\mathcal{R}(I) \neq \varnothing$ as shown in Fig. 11(b), which corresponds to actual connection processes in this system.

(2) **Interaction soundness**

Interaction soundness can be analyzed by using the corresponding algorithm, and the result is "the system has the interaction soundness". More specific steps are described as follows.

*Steps1-2.* Based on CT, final places $Fin_1$-$Fin_5$ of three components can be reached and receive tokens, such as the reachable configurations $\Pi_{f1}$ and $\Pi_{f2}$ of $N^m{}_{e2}$ in Fig. 11(a).

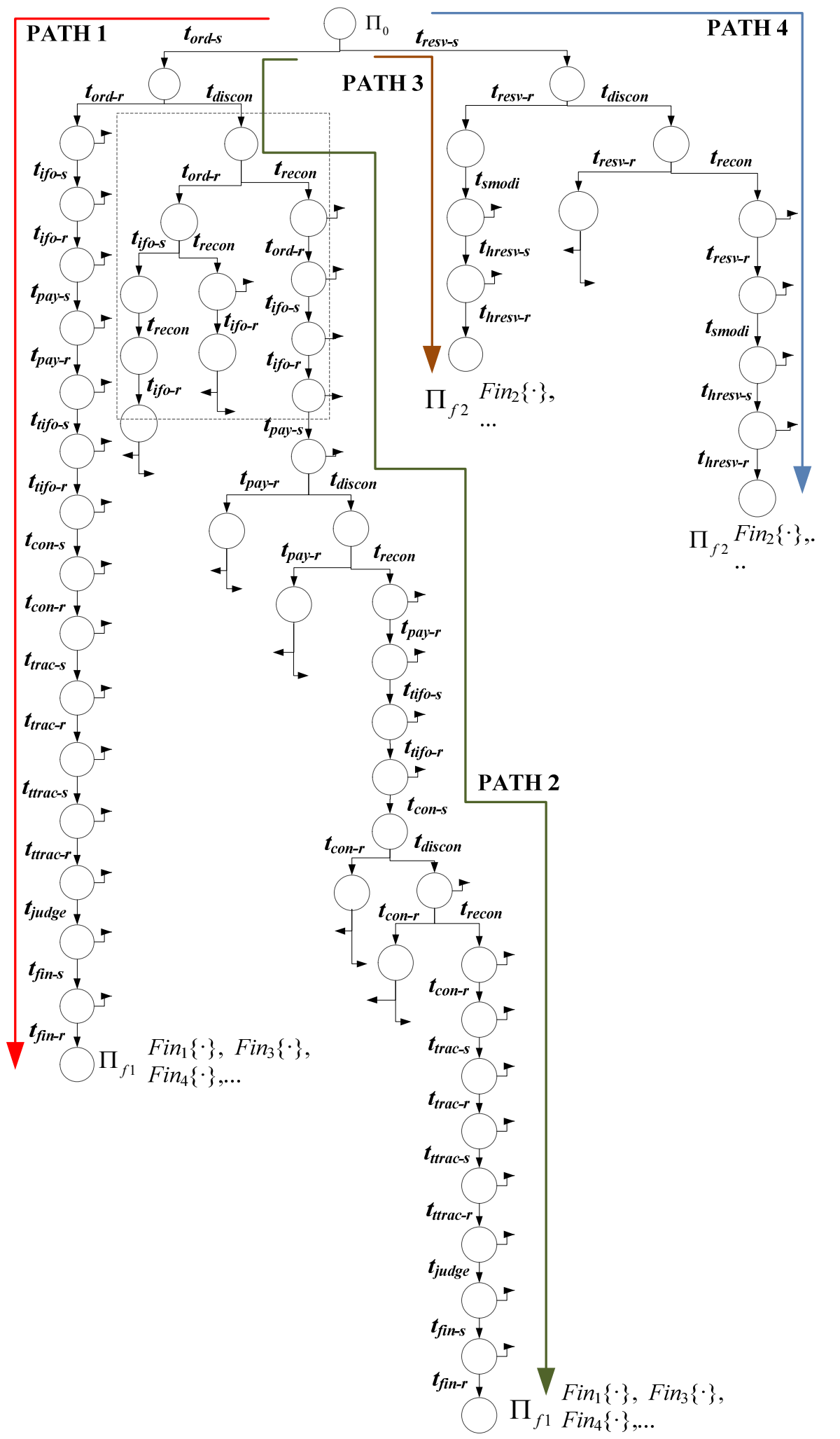

(a) A part of CT of $N^m{}_{e2}$

| | |
|---|---|
| $\mathcal{R}(I)$ | $placeorder_{C\text{-}M}$, $orderifo_{C\text{-}M}$, $transaction_{C\text{-}T}$, $transactionifo_{C\text{-}T}$, $transconfirm_{C\text{-}M}$, $transconfirm_{M\text{-}T}$, $confirmifo_{M\text{-}T}$, $final1_{C\text{-}M}$, $reservation_{C\text{-}M}$, $final2_{C\text{-}M}\}$ |
| $\mathcal{R}(TransID, OrderID, Gross, ReservIfo)$ | $transactionid$, $orderid$, $gross$, $reservationifo$ |
| $\mathbb{A}$ | $\{I \rightarrow placeorder_{C\text{-}M}, I \rightarrow transaction_{C\text{-}T}, I \rightarrow transconfirm_{C\text{-}M}, I \rightarrow transconfirm_{M\text{-}T}, I \rightarrow confirmifo_{M\text{-}T}, I \rightarrow reservation_{C\text{-}M}\}$ |
| $\mathbb{C}$ | $\{I \rightarrow placeorder_{C\text{-}M}, I \rightarrow orderifo_{C\text{-}M}, I \rightarrow transaction_{C\text{-}T}, I \rightarrow transactionifo_{C\text{-}T}, I \rightarrow transconfirm_{C\text{-}M}, I \rightarrow transconfirm_{M\text{-}T}, I \rightarrow confirmifo_{M\text{-}T}, I \rightarrow final1_{C\text{-}M}, I \rightarrow reservation_{C\text{-}M}, I \rightarrow final2_{C\text{-}M}\}$ |
| $\mathbb{K}$ | $\{I \rightarrow orderifo_{C\text{-}M}, I \rightarrow transactionifo_{C\text{-}T}, I \rightarrow final1_{C\text{-}M}, I \rightarrow final2_{C\text{-}M}\}$ |

| **Interfaces $\mathcal{I}$** | **CLI and MER ($f1$\|$f2$)** | | **CLI and TPP** | **MER and TPP** |
|---|---|---|---|---|
| **Connected interfaces** | $placeorder_{C\text{-}M}$, $orderifo_{C\text{-}M}$, $transconfirm_{C\text{-}M}$, $final1_{C\text{-}M}$, | $reservation_{C\text{-}M}$, $final2_{C\text{-}M}$ | $transaction_{C\text{-}T}$, $transactionifo_{C\text{-}T}$ | $transconfirm_{M\text{-}T}$, $confirmifo_{M\text{-}T}$ |
| **Disconnected Interfaces** | $orderifo_{C\text{-}M}$, $final1_{C\text{-}M}$ | $final2_{C\text{-}M}$ | $transactionifo_{C\text{-}T}$ | |

| | |
|---|---|
| **New-link sequences** | (1) $(I \rightarrow placeorder_{C\text{-}M}) \rightarrow (I \rightarrow orderifo_{C\text{-}M}) \rightarrow (I \rightarrow transaction_{C\text{-}T}) \rightarrow (I \rightarrow transactionifo_{C\text{-}T}) \rightarrow (I \rightarrow transconfirm_{C\text{-}M}) \rightarrow (I \rightarrow transconfirm_{M\text{-}T}) \rightarrow (I \rightarrow confirmifo_{M\text{-}T}) \rightarrow (I \rightarrow final1_{C\text{-}M})$<br>(2) $(I \rightarrow reservation_{C\text{-}M}) \rightarrow (I \rightarrow final2_{C\text{-}M})$ |
| **Broken-link sequences** | (1) $(I \rightarrow orderifo_{C\text{-}M}) \rightarrow (I \rightarrow transactionifo_{C\text{-}T}) \rightarrow (I \rightarrow final1_{C\text{-}M})$<br>(2) $(I \rightarrow final2_{C\text{-}M})$ |

(b) Some results of $N^m{}_{e2}$ based on its CT

Fig. 11. Analysis of VPN model $N^m{}_{e2}$ for Example 2.

*Step3.* We discover the change of $\gamma$ in this system, and obtain $\mathbb{A}$, $\mathbb{C}$ and $\mathbb{K}$ as shown in Fig. 11(b). It is noted that the above sets conform to the sets of possible connections and disconnections as shown in Figs. 7(a) and (b), and thus are not beyond the link capacity of the system.

*Steps4-5.* 1) Firstly we verify connections. As shown in Fig. 7, connections can happen among three components. The actual (connected) interface places instantiated from $I$ are obtained as shown in Fig. 11(b). According to the firing sequences in $N^m{}_{e2}$, they are all used to transfer different data by the firing of interaction transitions, such as $t_{ord\text{-}s}$, $t_{ord\text{-}r}$, $t_{ifo\text{-}s}$ and $t_{ifo\text{-}r}$, in Client, Merchant and TPP correspondingly.

2) Then we verify disconnections. In the system description, a disconnection may happen between Client and Merchant or TPP because of the mobility. Transition $t_{discon}$ represents disconnection while $t_{recon}$ means reconnection. According to CT, it can be found that firing $t_{discon}$ can disassociate $I$ with one of disconnected interfaces in Fig. 11(b), and one of receiving transitions in $\{t_{ifo\text{-}r}, t_{tifo\text{-}r}, t_{fin\text{-}r}, t_{hresv\text{-}r}\}$ from this interface cannot fire before firing $t_{recon}$. For example, after firing $t_{discon}$ with disassociating $I$ with $orderifo_{C\text{-}M}$, it is easy to conclude that $t_{ifo\text{-}r}$ does not fire before firing $t_{recon}$ and fires again after firing $t_{recon}$ in the execution (seen in the dashed box in Fig. 11(a)). Hence, disconnections are verified as expected.

Thus all interfaces in this system have good usability according to Definition 5.2.

Hence, interaction soundness of Example 2 has been verified. Contextual changes in this example are reflected by different instantiations of interfaces which have been analyzed. Here we do not elaborate them.

**(3) Data validity**

The data validity of $N^m{}_{e2}$ can be verified by the corresponding algorithm, and the result is also "**true**". The detailed process is introduced as follows.

*Step1.* Data synchronization (matching) exists in the firings of three transitions $t_{judge}$, $t_{smodi}$ and $t_{ttrac\text{-}s}$ in *MER* and *TPP*. It can be easily analyzed that the input arcs of each of them can all match when firing each of them according to Definition 3.6. We explain the data synchronization in $N^m{}_{e2}$ as an example. When Merchant receives the transaction confirmation of Client, it sends the corresponding transaction information to TPP. Then TPP should match the transaction information (*TransID*) received from Merchant with its storage ($t_{ttrac\text{-}s}$) and return the corresponding order information to Merchant. Merchant receives the order information, and matches it with its storage, modifies the order state ($t_{judge}$), and returns the result to Client finally. Similar to $t_{judge}$, $t_{smodi}$ is to modify the state of a reservation. According to binding function $\mathcal{V}$ of transitions and the net execution in CT, the above parameters are all matched, and $t_{judge}$, $t_{smodi}$ and $t_{ttrac\text{-}s}$ can fire as usual.

*Step2.* It should be noted that if the data in any storage place in $\{OrdI, OInf, RInf, State1, State2\}$ of *MER* or *TPP* has been tampered with, the data synchronization fails ($t_{judge}$ or $t_{smodi}$ or $t_{ttrac\text{-}s}$ cannot fire) and a deadlock occurs. The data synchronization is important in this system, and places *OrdI*, *OInf*, *RInf*, *State1* and *State2* are vulnerable and critical, which have been denoted by grey circles in Fig. 10(a).

*Step3.* Based on the binding sequences in the model, the formal parameters *TransID*, *OrderID*, *Gross* and *ReservIfo* (which are used for the data) only have one instantiation in the interfaces in the net execution (Fig. 10(b)).

*Step4.* It can be easily found that each actual interface instantiated by $I$ can have at most one token in the consecutive net execution and every token in them can be transferred by firing an interaction transition.

According to the above discussion, properties of the system are represented and analyzed based on VPN. It is noted that the designed system can run normally, and have system connectivity, interaction soundness and data validity. However, the analysis also indicates that some places (locations) and transitions (actions) in VPN may be vulnerable and critical, and need more attention.

## VII. Conclusion

This work studies the modeling and analysis of mobile interactive systems based on a newly proposed Variable Petri Nets (VPN). Firstly we give the description of mobile interactive systems. Then we introduce the definition and firing rule of VPN. VPN owns a dynamic structure that can be used to model uncertain interactions in mobile interactive systems. The analysis techniques for VPN are also presented. Then we propose a VPN model construction method for mobile interactive systems. Based on the obtained model, we introduce three critical properties about interactions which need to be considered in the system design, and their related analysis methods. Finally, we use a practical example to illustrate the proposed concepts and methods. Our new method is useful in describing and verifying some important properties of mobile interactive systems.

The resulting model is complicated for even a small system. Hence in the future research, we intend to focus on the simplification of the proposed model as well as more specific and practical analysis techniques especially some algebraic analysis approaches to mobile interactive systems.

## References

[1] M. Weiser. The computer for the 21st century. *Scientific American*. 1991; 265(3):66.

[2] G. W. Musumba, H. O. Nyongesa. Context awareness in mobile computing: A review. *International Journal of Machine Learning & Applications*, 2013, 2(1).

[3] J. Ye, S. Dobson, S. Mckeever. *Situation identification techniques in pervasive computing: A review*. Elsevier Science Publishers B. V. 2012.

[4] C. Perera, A. Zaslavsky, P. Christen, et al. Context Aware Computing for The Internet of Things: A Survey. *IEEE Communications Surveys & Tutorials*, 2013, 16(1):414-454.

[5] A. U. R. Khan, M. Othman, S. A. Madani, et al. A Survey of Mobile Cloud Computing Application Models. *IEEE Communications Surveys & Tutorials*, 2014, 16(1):393-413.

[6] S. Deng, L. Huang, J. Taheri, J. Yin, M. C. Zhou, and A. Y. Zomaya, “Mobility-Aware Service Composition in Mobile Communities,” *IEEE Transactions on Systems, Man, and Cybernetics: Systems*, Vol. 47, No. 3, pp. 555 – 568, Mar. 2017.

[7] J. L. Peterson, *Petri Net Theory and the Modeling of Systems*. Englewood Cliffs, NJ: Prentice-Hall, 1981.

[8] T. Murata. Petri nets: Properties, analysis and applications. *Proceedings of the IEEE*, 1989, 77(4):541-580.

[9] K. Jensen, L. M. Kristensen, Coloured Petri nets: a graphical language for formal modeling and validation of concurrent systems. *Communications of the ACM*, 2015, 58(6): 61-70.

[10] T. Murata, D. Zhang. A predicate-transition net model for parallel interpretation of logic programs. *IEEE Transactions on Software Engineering*, 1988, 14(4):481-497.

[11] R. Wang, S. Chen, X. F.Wang, S. Qadeer. How to Shop for Free Online -- Security Analysis of Cashier-as-a-Service Based Web Stores. *IEEE Symposium on Security and Privacy*. IEEE Computer Society, 2011:465-480.

[12] W. Y. Yu, C. G. Yan, Z. J. Ding, et al. Modeling and Validating E-Commerce Business Process Based on Petri Nets. *IEEE Transactions on Systems Man & Cybernetics Systems*, 2014, 44(3):327-341.

[13] M. Elkoutbi, R. K. Keller, Modeling interactive systems with hierarchical Coloured petri nets. *Advanced Simulation Technologies Conference*. 1998: 432-37.

[14] H. Ezzedine, C. Kolski. Use of Petri Nets for Modeling an Agent-Based Interactive System: Basic Principles and Case Study. *Petri Net Theory & Applications*, 2008:131-148.

[15] G. Liu, C. Jiang, M. Zhou, P. Xiong, Interactive petri nets. *IEEE Transactions on Systems, Man, and Cybernetics: Systems*, 2013, 43(2), 291-302.

[16] O. O. Captarencu. Modelling and Verification of Interorganizational Workflows with Security Constraints: A Petri Nets-Based Approach. *Lecture Notes in Business Information Processing*, 2012, 112:486-493.

[17] M. P. Van Der Aalst. Modeling and analyzing interorganizational workflows. *International Conference on Application of Concurrency To System Design. Proceedings.* IEEE, 1998: 262-272.

[18] A. P. Estrada-Vargas, E. López-Mellado, J. J. Lesage. Automated modelling of reactive discrete event systems from external behavioural data. *International Conference on Electronics, Communications and Computing.* IEEE, 2013:120-125.

[19] R. Valk, Concurrency in communicating object Petri nets. *Concurrent Object-Oriented Programming and Petri Nets*. Springer Berlin Heidelberg, 2001: 164-195.

[20] L. W. Dworzanski, I. A. Lomazova. Structural Place Invariants for Analyzing the Behavioral Properties of Nested Petri Nets. *Application and Theory of Petri Nets and Concurrency*. Springer International Publishing, 2016: 325-344.

[21] L. Chang, X. He, J. Lian, and S. Shatz: "Applying a Nested Petri Net Modeling Paradigm to Coordination of Sensor Networks with Mobile Agents", *Proceeding of Workshop on Petri Nets and Distributed Systems 2008*, Xian, China, June, 2008, pp.132-145.

[22] Y. Kissoum, R. Maamri, Z. Sahnoun. Modeling smart home using the paradigm of nets within nets. *International Conference on Artificial Intelligence: Methodology, Systems, and Applications*. Springer-Verlag, 2012:286-295.

[23] F. Cristini, C. Tessier. Nets-within-Nets to Model Innovative Space System Architectures. *International Conference on Application and Theory of Petri Nets*. Springer-Verlag, 2012:348-367.

[24] M. Llorens, J. Oliver. Structural and dynamic changes in concurrent systems: reconfigurable Petri nets. *IEEE Transactions on Computers*, 2004, 53(9): 1147-1158.

[25] L. Kahloul, A. Chaoui, K. Djouani, Modelling and Analysis of Mobile Computing Systems: An Extended Petri Nets Formalism. *International Journal of Computers Communications & Control*, 2015, 10(2): 211-221.

[26] L. Kahloul, S. Bourekkache, K. Djouani. Designing reconfigurable manufacturing systems using reconfigurable object Petri nets. *International Journal of Computer Integrated Manufacturing*, 2016:1-18.

[27] D. X. Xu, Y. Deng, J. H. Ding, A formal architectural model for logical agent mobility. *IEEE Transactions on Software Engineering*, 2003, 29(1):31-45.

[28] T. Miyamoto, K. Horiguchi. Modular Reachability Analysis of Petri Nets for Multiagent Systems. *IEEE Transactions on Systems, Man & Cybernetics Systems*, 2013, 43(6):1411-1423.

[29] G. Yasuda. Behavior-based autonomous cooperative control of intelligent mobile robot systems with embedded Petri nets. *International Symposium on Soft Computing and Intelligent Systems. IEEE*, 2014:1085-1090.

[30] S. L. Jin, M. C. Zhou, P. L. Hsu. A Petri-Net Approach to Modular Supervision With Conflict Resolution for Semiconductor Manufacturing Systems. *IEEE Transactions on Automation Science & Engineering*, 2007, 4(4):584-588.

[31] M. Kloetzer, C. Mahulea. Accomplish multi-robot tasks via Petri net models. *IEEE International Conference on Automation Science and Engineering*. IEEE, 2015:304-309.

[32] R. Kodikara, S. Ling, Zaslavsky A. Evaluating Cross-layer Context Exchange in Mobile Ad-hoc Networks with Colored Petri Nets. *IEEE International Conference on Pervasive Services*. IEEE, 2007:173-176.

[33] M. H. Ghahramani, M. C. Zhou, and C. T. Hon, "Toward Cloud Computing QoS Architecture: Analysis of Cloud Systems and Cloud Services," *IEEE/CAA Journal of Automatica Sinica*, Vol. 4, No. 1, pp. 5-17, Jan. 2017.
[34] P. Zhang, M. Zhou and G. Fortino, "Security and trust issues in Fog computing: A survey", *Future Generation Computer Systems*, Vol. 88, pp. 16-27, 2018.
[35] Y. Guo, X. Hu, B. HU, J. Cheng, M. Zhou and R. Y. K. Kwok, "Mobile Cyber Physical Systems: Current Challenges and Future Networking Applications," IEEE Access, Vol. 6, pp. 12360-12368, 2018.
[36] X. Lu, M. Zhou, A. C. Ammari, and J. Ji, "Hybrid Petri Nets for Modeling and Analysis of Microgrid Systems," *IEEE/CAA Journal of Automatica Sinica*, 3(4), pp. 347-354, Oct. 2016.
[37] F. J. Yang, N. Q. Wu, Y. Qiao, and R. Su, "Polynomial approach to optimal one-wafer cyclic scheduling of treelike hybrid multi-cluster tools via Petri nets," *IEEE/CAA J. of Autom. Sinica*, vol. 5, no. 1, pp. 270-280, Jan. 2018.
[38] J. J. Cheng, C. Liu, et al. "Automatic composition of semantic web services based on fuzzy predicate petri nets." *IEEE Transactions on Automation Science and Engineering* 12.2 (2015): 680-689.
[39] H. Chen, L. Amodeo, F. Chu, et al. Modeling and performance evaluation of supply chains using batch deterministic and stochastic Petri nets. *IEEE Transactions on Automation Science and Engineering*, 2005, 2(2): 132-144.
[40] "Variable Petri nets". *IEEE Transactions on Systems Man & Cybernetics Systems*. submitted. 2020

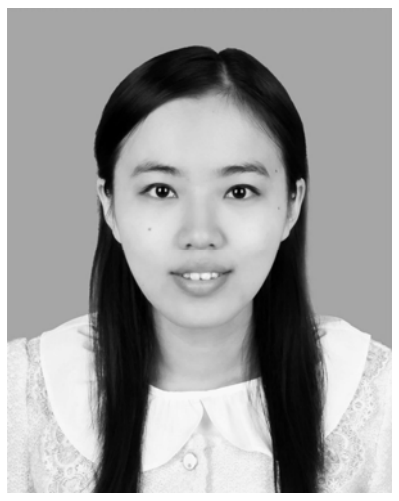

**Ru Yang** received the B.S. degree from Shandong University of Science and Technology, Qingdao, China, in 2013. She is currently pursuing the Ph.D. degree with the Department of Computer Science and Technology, Tongji University, Shanghai, China.

Her current research interests include Petri nets and formal engineering.

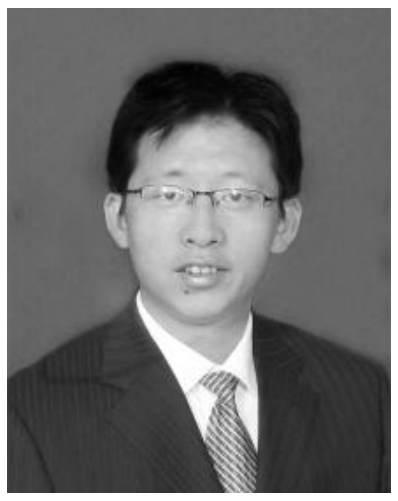

**ZhiJun Ding** received the M.S. degree from Shandong University of Science and Technology, Taian, China, in 2001, and Ph.D. degree from Tongji University, Shanghai, China, in 2007.

Now he is a Professor of the Department of Computer Science and Technology, Tongji University. His research interests are in formal engineering, Petri nets, services computing, and workflows. He has published more than 100 papers in domestic and international academic journals and conference proceedings.

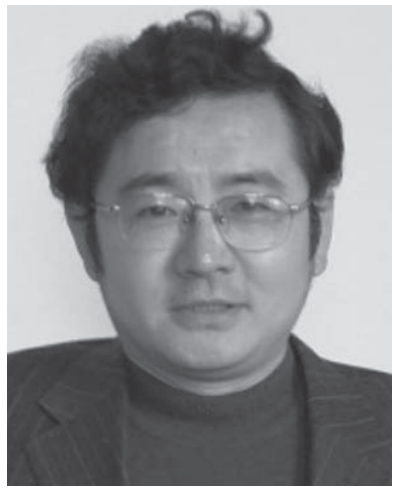

**ChangJun Jiang** received the Ph.D. degree from the Institute of Automation, Chinese Academy of Sciences, Beijing, China, in 1995.

He is currently a Professor with the Department of Computer Science and Technology, Tongji University, Shanghai, China. His current research interests include concurrency theory, Petri nets, formal verification of software, cluster, grid technology, program testing, intelligent transportation systems, and service-oriented computing. He has published more than 100 publications.

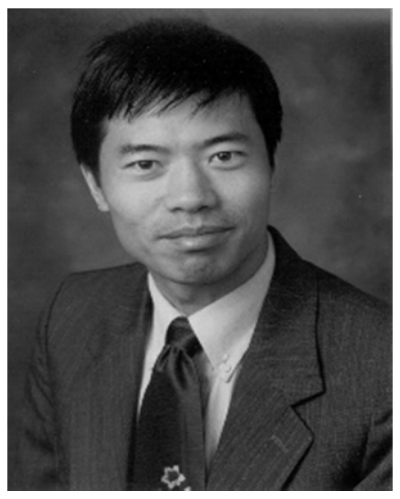

**MengChu Zhou** (S'88-M'90-SM'93-F'03) received his B.S. degree in Electrical Engineering from Nanjing University of Science and Technology, Nanjing, China in 1983, M.S. degree in Automatic Control from Beijing Institute of Technology, Beijing, China in 1986, and Ph. D. degree in Computer and Systems Engineering from Rensselaer Polytechnic Institute, Troy, NY in 1990.

He joined the New Jersey Institute of Technology, Newark, NJ, USA, in 1990, and became a Distinguished Professor of Electrical and Computer Engineering in 2013. His current research interests include Petri nets, sensor networks, Web services, semiconductor manufacturing, transportation, and energy systems. He has over 600 publications including 12 books, 200+ journal papers (majority in IEEE TRANSACTIONS), and 28 book chapters.

Dr. Zhou is the Founding Editor of the IEEE Press Book Series on Systems Science and Engineering. He is a Life Member of the Chinese Association for Science and Technology—USA and served as its President in 1999. He is a fellow of the American Association for the Advancement of Science.